\documentclass[twocolumn]{ECMS}
\usepackage{preamble}

\author{
    \normalsize
    \begin{tabular}{cc}
        Vyacheslav Zhdanovskiy & Lev Teplyakov \\
        Institute for Information & Institute for Information \\
        Transmission Problems, RAS & Transmission Problems, RAS \\ 
        Bolshoy Karetny per.~19, Moscow, 127051, Russia; & Bolshoy Karetny per.~19, Moscow, 127051, Russia \\
        Moscow Institute of Physics and Technology & E-mail: teplyakov@visillect.com \\
        (National Research University) & \\
        Institutskiy per.~9, Dolgoprudny, 141701, Russia & \\
        E-mail: zhdanovskiy.vd@phystech.edu & \\
        & \\
        \multicolumn{2}{c}{Anton Grigoryev} \\
        \multicolumn{2}{c}{Institute for Information Transmission Problems, RAS} \\ 
        \multicolumn{2}{c}{Bolshoy Karetny per.~19, Moscow, 127051, Russia} \\
        \multicolumn{2}{c}{E-mail: me@ansgri.com} \\ 
    \end{tabular}
}

\title{\textbf{PREDICTING PERFORMANCE OF HETEROGENEOUS AI SYSTEMS WITH DISCRETE-EVENT SIMULATIONS}}
\begin{document}
\voffset=-0.5in
\maketitle

\copyrightnote{
	\begin{tabular}{l}
		Communications of the ECMS, Volume 36, Issue 1, Proceedings, \\
		©ECMS Ibrahim A. Hameed, Agus Hasan, Saleh Abdel-Afou Alaliyat (Editors) \\
		ISBN: 978-3-937436-77-7/978-3-937436-76-0(CD) ISSN 2522-2414
	\end{tabular}
}

\thispagestyle{empty}\pagestyle{empty}



\section*{\textbf{KEYWORDS}}
Heterogeneous computing; Discrete-event simulations; Artificial intelligence; Video analytics; Software architecture

\section*{\textbf{ABSTRACT}}
In recent years, artificial intelligence (AI) technologies have found industrial applications in various fields.
AI systems typically possess complex software and heterogeneous CPU/GPU hardware architecture, making it difficult to answer basic questions considering performance evaluation and software optimization.
Where is the bottleneck impeding the system? How does the performance scale with the workload?
How the speed-up of a specific module would contribute to the whole system?
Finding the answers to these questions through experiments on the real system could require a lot of computational, human, financial, and time resources.
A solution to cut these costs is to use a fast and accurate simulation model preparatory to implementing anything in the real system.
In this paper, we propose a discrete-event simulation model of a high-load heterogeneous AI system in the context of video analytics.
Using the proposed model, we estimate:
1) the performance scalability with the increasing number of cameras;
2) the performance impact of integrating a new module;
3) the performance gain from optimizing a single module.
We show that the performance estimation accuracy of the proposed model is higher than 90\%.
We also demonstrate, that the considered system possesses a counter-intuitive relationship between workload and performance, which nevertheless is correctly inferred by the proposed simulation model.
\section*{\textbf{INTRODUCTION}}
In recent years, there has been a significant growth of interest in machine learning and artificial intelligence (AI) technologies.
Analysts expect the AI market to grow to more than USD 660 billion by 2028 \citep{aimarketreport}.
Nowadays, AI technologies are used in various fields, such as urban services \citep{wang2015deep}, retail \citep{weber2019state}, medicine \citep{chen2021survey}, etc.

One of the key technologies that allowed for the progress is deep learning.
In particular, deep neural networks made it possible to achieve almost human-like object recognition quality in problems like image classification \citep{rawat2017deep}, object detection \citep{arnold2019survey, liu2020deep} and image segmentation \citep{guo2018review}.
In order to achieve this accuracy and still provide reasonable recognition speed, deep neural networks have to exploit special hardware providing fast matrix multiplication, most commonly --- GPUs.
This poses a problem for software developers, because they have to design the architecture of their AI-based applications with heterogeneous CPU/GPU computations in mind.

In detail, software developers have to utilize the parallelism provided by modern multi-core CPUs along with the GPU acceleration.
Meanwhile the GPU is used for the neural network inference, the CPU is used to prepare the neural networks's input and postprocess its output, run various computer vision algorithms like tracking, localization, keypoint detection.
The CPU also handles all the additional modules delivering AI results to the end user --- API calls, business logic, database management, etc.
In order to deliver on all these tasks on advanced heterogeneous hardware with the given time, memory and energy constraints, software developers have to design rather complex architectures \citep{sutter2005free}.

The complex software architecture, in turn, complicates performance evaluation and software optimization of such systems \citep{Voss2019flowgraphbeyond}.
In particular, it is hard to locate the bottleneck module and to predict, how its optimization will affect the performance of the entire system.
It may lead to lots of human and financial resources as well as time resources being spent on optimization without any significant increase in performance of the entire system.
For a new module to be designed and implemented, it might be hard to map the system's constraints to the constraints of a specific module.
It may turn out - after the resources are spent on implementing a new module and its integration into the system! - that the module results in unaffordable slowdown of the system.

An elegant way to avoid the aforementioned problems is to design a simulation model of the system and infer the feasibility of optimizations on the model prior to implementing them in the code.
Discrete-event simulations of software have been used before in other areas, for example, planning, evaluation and optimization of Hadoop clusters \citep{bian2014simulating, wang2014simulating, liu2016planning, chen2016cluster}.
However, to the best of our knowledge, no one yet has tried to apply this approach to applications with heterogeneous CPU/GPU computing, such as AI systems.
Moreover, such a simulation model can be used for other tasks like capacity planning.
This can be achieved by evaluating the model on a collected set of suitable hardware configurations \citep{korobov2020swapc}.

In terms of performance modelling, different AI applications have their own unique specialties.
The key difference is which hardware components to consider.
For example, video analytics \citep{wang2015deep} and self-driving cars \citep{badue2021self} are mostly CPU/GPU intensive, meanwhile AI on pure-cloud solutions also rely heavily on the network performance, which requires to consider the I/O subsystem in the model.

In this work, we consider a video analytics system as a typical example of AI application utilizing CPU/GPU computing.
The possibility of applying our research to other AI applications is discussed in Section~III.
While the GPU is used to infer the neural network responsible for complex object detection tasks such as human detection, the CPU is used for preprocessing the video frames, postprocessing the neural network output and running classic computer vision algorithms such as ORB keypoint detector and descriptor \citep{rublee2011orb}.
The considered system has the following functionality:
\begin{itemize}
    \item processing video feed from multiple video cameras;
    \item person detection using YOLOv4 \citep{bochkovskiy2020yolov4};
    \item person 3D localization;
    \item single- and multi-camera person tracking.
\end{itemize}

We design a discrete-event simulation model which is low-cost both in terms of its development and simulation speed and can easily be adopted by the software developers.
We use it to estimate:
\begin{enumerate}
    \item the performance scalability with the increasing number of video cameras;
    \item the performance gain from optimizing a single system module;
    \item the performance impact caused by integrating of a new module in the system.
\end{enumerate}

The rest of the paper is structured as follows: Section~I describes the software architecture of the video analytics system, Section~II provides a description of the proposed simulation model, Section~III contains numerical results. Section~IV concludes the paper.

\section*{\textbf{SYSTEM DESCRIPTION}}
\subsection*{\textbf{Flow graph paradigm}}
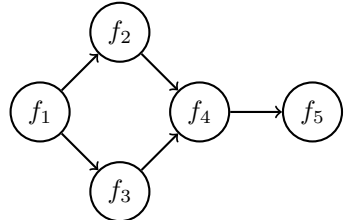
\begin{figure}[h!]
    \centering
    \begin{tikzpicture}[node distance={15mm}, thick, main/.style = {draw, circle}] 
        \node[main] (1) {$f_1$}; 
        \node[main] (2) [above right of=1] {$f_2$}; 
        \node[main] (3) [below right of=1] {$f_3$}; 
        \node[main] (4) [above right of=3] {$f_4$}; 
        \node[main] (5) [right of=4] {$f_5$}; 
        \draw[->] (1) -- (2); 
        \draw[->] (1) -- (3); 
        \draw[->] (2) -- (4); 
        \draw[->] (3) -- (4); 
        \draw[->] (4) -- (5); 
    \end{tikzpicture} 
    \caption{Example of a flow graph. Nodes $f_2$ and $f_3$ can process messages broadcasted from $f_1$ in parallel. $f4$ can process messages from $f_2$ and $f_3$ independently or wait until messages from both nodes are present --- it depends on the desired behaviour.}
    \label{fig:flow_graph}
\end{figure}

A common design pattern to efficiently implement parallelism in heterogeneous and parallel systems is to use the flow graph paradigm \citep{grigoryev2015building, badue2021self, huang2021taskflow}. 
With it, the algorithm is represented as a data flow graph \citep{Voss2019flowgraph}, see Fig.~\ref{fig:flow_graph}.
Each graph node (vertex) receives a message from its predecessors, processes it and broadcasts the output to the successors.
Input and output nodes are the exception:
\begin{itemize}
    \item input nodes either produce an input message by itself or receive it somewhere from outside the graph;
    \item output nodes do not broadcast the result, but instead store it or deliver it outside the graph.
\end{itemize}

The considered video analytics system has multiple places where parallelism can be utilized. In particular:
\begin{itemize}
    \item frames from different video streams can be processed in parallel;
    \item a single frame can be processed in parallel by multiple data-independent detectors, for example, by the neural network and the keypoint detector.
\end{itemize}

Flow graphs allow to efficiently utilize these types of parallelism by executing different nodes of the graph in parallel at the same time.

There are multiple frameworks implementing this paradigm. Notable examples include oneTBB\footnote{Formerly TBB; more details at: \url{https://www.intel.com/content/www/us/en/developer/tools/oneapi/onetbb.html}} and Taskflow\footnote{More details at: \url{https://taskflow.github.io/}} \citep{huang2021taskflow}.
In this work, we use oneTBB.

\subsection*{\textbf{Software architecture of the video analytics system}}
\begin{figure*}
    \centering
    \begin{subfigure}{1\textwidth}
        \centering
        \begin{tikzpicture}[
                node distance=0.5cm,
                stream1/.style = {draw, rectangle, align=center, font=\small, rounded corners=.2cm, fill=blue!10!white, minimum height=1.0cm},
                stream2/.style = {draw, rectangle, align=center, font=\small, rounded corners=.2cm, fill=red!10!white, minimum height=1.0cm},
                gpu/.style = {draw, rectangle, align=center, font=\small, rounded corners=.2cm, fill=yellow!10!white, minimum height=1.25cm}
            ] 
            \node[stream1] (1_preprocessor) {Node A};
            
            \node[gpu] (nn_inferencer) [below right=of 1_preprocessor] {GPU Node};
            
            \node[stream1] (1_nn_postprocessor) [above right=of nn_inferencer] {Node B};
            \node[stream1] (1_tracking) [right=of 1_nn_postprocessor] {Node C};
            \node[stream1] (1_output) [right=of 1_tracking] {Node D};
            \node[stream1] (1_cpu_detector) [above right=of 1_preprocessor] {Node E};
            
            \draw[->] (1_preprocessor.east) -- (nn_inferencer.north);
            \draw[->] (nn_inferencer.north) -- (1_nn_postprocessor.west);
            \draw[->] (1_nn_postprocessor.east) -- (1_tracking.west);
            \draw[->] (1_tracking.east) -- (1_output.west);
            \draw[->] (1_preprocessor.north) -- (1_preprocessor|-1_cpu_detector) -- (1_cpu_detector.west);
            \draw[->] (1_cpu_detector.east) -- (1_cpu_detector-|1_output) -- (1_output.north);
            
            \node[stream2] (2_preprocessor) [below left=of nn_inferencer] {Node A};
            \node[stream2] (2_nn_postprocessor) [below right=of nn_inferencer] {Node B};
            \node[stream2] (2_tracking) [right=of 2_nn_postprocessor] {Node C};
            \node[stream2] (2_output) [right=of 2_tracking] {Node D};
            \node[stream2] (2_cpu_detector) [below right=of 2_preprocessor] {Node E};
            
            \draw[->] (2_preprocessor.east) -- (nn_inferencer.south);
            \draw[->] (nn_inferencer.south) -- (2_nn_postprocessor.west);
            \draw[->] (2_nn_postprocessor.east) -- (2_tracking.west);
            \draw[->] (2_tracking.east) -- (2_output.west);
            \draw[->] (2_preprocessor.south) -- (2_preprocessor|-2_cpu_detector) -- (2_cpu_detector.west);
            \draw[->] (2_cpu_detector.east) -- (2_cpu_detector-|2_output) -- (2_output.south);
            
            \node (1_label) [left=0.5 cm of 1_preprocessor, align=center] {\footnotesize Video \\ stream \#1};
            \node (2_label) [left=0.5 cm of 2_preprocessor, align=center] {\footnotesize Video \\ stream \#2};
            \node (3_label) [right=0.5 cm of 1_output, align=center] {\footnotesize Output};
            \node (4_label) [right=0.5 cm of 2_output, align=center] {\footnotesize Output};
            \draw[->] (1_label.east) -- (1_preprocessor.west);
            \draw[->] (2_label.east) -- (2_preprocessor.west);
            \draw[->] (1_output.east) -- (3_label.west);
            \draw[->] (2_output.east) -- (4_label.west);
        \end{tikzpicture} 
        \caption{}
        \label{fig:vas_graph}
        \vspace{5pt}
    \end{subfigure}
    \begin{subfigure}{1\textwidth}
        \centering
        \begin{tikzpicture}[
                node distance=0.0cm,
                stream1/.style = {draw, rectangle, align=center, font=\small, fill=blue!10!white, minimum height=0.9cm, minimum width=2cm},
                stream2/.style = {draw, rectangle, align=center, font=\small, fill=red!10!white, minimum height=0.9cm, minimum width=2cm},
                gpu/.style = {draw, rectangle, align=center, font=\small, fill=yellow!10!white, minimum height=0.9cm}
            ] 
            \node[stream1] (1_preprocessor) {Node A};
            \node[stream2] (2_preprocessor) [below=0.1cm of 1_preprocessor] {Node A};
            \node[stream1] (1_module) [right=of 2_preprocessor] {Node E};
            \node[stream2] (2_module) [above=0.1cm of 1_module] {Node E};
            
            \node[gpu] (1_nn_inferencer) [below right=0.1 cm and 0cm of 2_preprocessor, minimum width=2.5cm] {GPU Node};
            \node[stream1] (1_nn_postprocessor) [above right= 1.1 cm and 0 cm of 1_nn_inferencer] {Node B};
            \node[stream1] (1_tracking) [below right=0.1cm and 0 cm of 1_nn_postprocessor, minimum width=4cm]  {Node C};
            \node[stream1] (1_output)       [right=of 1_tracking]      {Node D};
            \node[gpu] (2_nn_inferencer) [right=of 1_nn_inferencer, minimum width=1.9cm] {GPU Node};
            \node[stream2] (2_nn_postprocessor) [above right=1.1 cm and 0 cm of 2_nn_inferencer] {Node B};
            \node[stream2] (2_tracking) [right=of 2_nn_postprocessor, minimum width=2cm]  {Node C};
            \node[stream2] (2_output)       [right=of 2_tracking]      {Node D};
            
            \node (1_label) [left=0.1 cm of 1_preprocessor, align=center] {CPU core \#1};
            \node (2_label) [left=0.1 cm of 2_preprocessor, align=center] {CPU core \#2};
            \node (3_label) at (1_nn_inferencer -| 2_label) {GPU};
            
            \node (1_cont) [right=0.5 cm of 2_output, minimum height=0.9cm] {\large ...};
            \node (2_cont) [below=0.1 cm of 1_cont, minimum height=0.9cm] {\large ...};
            \node (3_cont) [below=0.1 cm of 2_cont, minimum height=0.9cm] {\large ...};
            
            \draw[->] ($(1_preprocessor.west)-(0,2.5)$)--($(1_preprocessor.west)-(-13.9,2.5)$) node [anchor=north] {time};
        \end{tikzpicture} 
        \caption{}
        \label{fig:timing}
        \vspace{5pt}
    \end{subfigure}
    \caption{Software architecture of the studied video analytics system represented as a flow graph (a). Sample timing diagram of the system execution (b). Note than \texttt{Node E} is processed in parallel with the \texttt{GPU Node}, meanwhile \texttt{Node B} has to wait until \texttt{GPU Node} finishes processing its task.
    \texttt{Node E} is the CPU-intensive module we consider in Section~III. It is turned off for all experiments except ``Integration of a new module in the system''.}
\end{figure*}
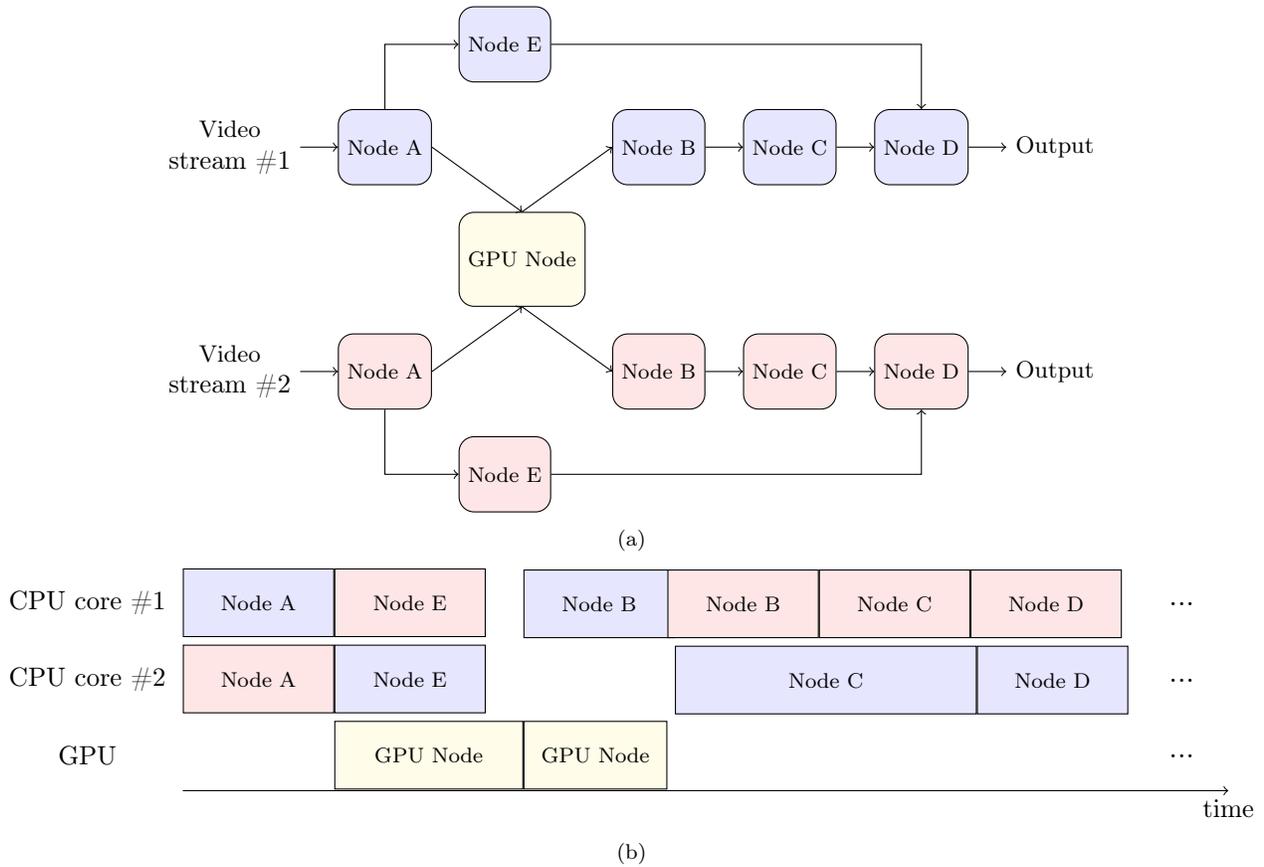
In this section, we provide a brief description of the software architecture used in the considered video analytics system.

Overall, the whole system is represented by a flow graph, see Fig.~\ref{fig:vas_graph}.
Each video stream from a single camera is represented by a separate graph component.
We shall notice that we deliberately show a less detailed version of the real graph for the sake of simplification.
All graph components are executed in a single thread pool.

Meanwhile many flow graph nodes can be executed in parallel, some of them can not because their functions are not thread safe.
In particular, this is true with the neural network nodes.
Because neural networks are usually implemented in a way \citep{tensorrtdocs} that each neural network instance can be executed on GPU by only one context at a particular moment, multiple video streams have to put their tasks for the neural network inference in a shared queue and then wait for their completion, see Fig.~\ref{fig:timing}.
We implement this functionality with oneTBB's \textit{async nodes} \citep{Voss2019async}.
\section*{\textbf{SIMULATION MODEL}}
In this section, we describe the details of the proposed simulation model.
Overall, the system flow graph has a near-direct representation in the model.
We implemented the model in Python using the SimPy\footnote{More details at: \url{https://simpy.readthedocs.io/en/latest/index.html}} library.
\begin{algorithm}[h]
    \caption{Flow graph node simulation for a basic node}\label{alg:basic}
    \begin{algorithmic}
        \State $Q$ --- \texttt{input queue (from predecessors)}
        \State $S$ --- \texttt{output queue (to successors)}
        \State $C$ --- \texttt{pool of free CPU cores}
        \State $P$ --- \texttt{distribution of the node's running time}
        \While{\texttt{not all frames are processed}}
            \State $m \gets $ \texttt{Q.pop()} \texttt{(blocks if $Q$ is empty)}
            \State $c \gets $ \texttt{C.pop()} \texttt{(blocks if $C$ is empty)}
            \State $t \sim P$
            \State \texttt{wait}$(t)$
            \State \texttt{$S$.push($m$)}
            \State \texttt{C.push($c$)}
        \EndWhile
    \end{algorithmic}
\end{algorithm}

\begin{algorithm}
    \caption{Flow graph node simulation for a GPU async node}\label{alg:async}
    \begin{algorithmic}
        \State $Q$ --- \texttt{input queue (from predecessors)}
        \State $S$ --- \texttt{output queue (to successors)}
        \State $C$ --- \texttt{pool of free CPU cores}
        \State $G$ --- \texttt{GPU}
        \State $P$ --- \texttt{distribution of the node's running time}
        \While{\texttt{not all frames are processed}}
            \State $m \gets $ \texttt{Q.pop()} \texttt{(blocks if $Q$ is empty)}
            \State $c \gets $ \texttt{C.pop()} \texttt{(blocks if $C$ is empty)}
            \State \texttt{G.lock()} \texttt{(blocks if $G$ is busy)}
            \State \texttt{C.push(c)}
            \State $t \sim P$
            \State \texttt{wait}$(t)$
            \State \texttt{G.release()}
            \State $c \gets $ \texttt{C.pop()} \texttt{(blocks if $C$ is empty)}
            \State \texttt{$S$.push($m$)}
            \State \texttt{C.push($c$)}
        \EndWhile
    \end{algorithmic}
\end{algorithm}

The source code of the implementation is available at GitHub\footnote{More details at: \url{https://github.com/iitpvisionlab/heterogeneous-ai-system-simulator}}.
It contains only $\approx400$ LoC of Python (including the profiling trace parsers), meanwhile the real system contains $\approx15000$ LoC of C++ (not including the dependencies).

We consider a CPU with $N$ logical cores in the model.
Each flow graph node waits until the input message is present, see Algorithm~\ref{alg:basic}. 
Then it waits until there is a free CPU core, then the node occupies the CPU core for the execution time. 
The execution time is sampled from independent empirical distributions measured by profiling the real system. 
Input messages that have not yet been processed are stored in a queue.
Effectively, this represents the work of oneTBB's thread pool.

The GPU neural network node is modelled in a slightly different way, see Algorithm~\ref{alg:async}.
Like a basic flow graph node, it waits for an input message and a CPU core.
Then it waits until the GPU is free, then locks it for the execution time, while yielding the CPU core for some another flow graph node.
When the GPU is done with processing the message, the node waits again for a free CPU core in order to broadcast its output to the successors.
Effectively, this represents the mechanism shown in Fig.~\ref{fig:timing}.

\begin{figure}[t!]
    \centering
    \includegraphics[width=1\linewidth]{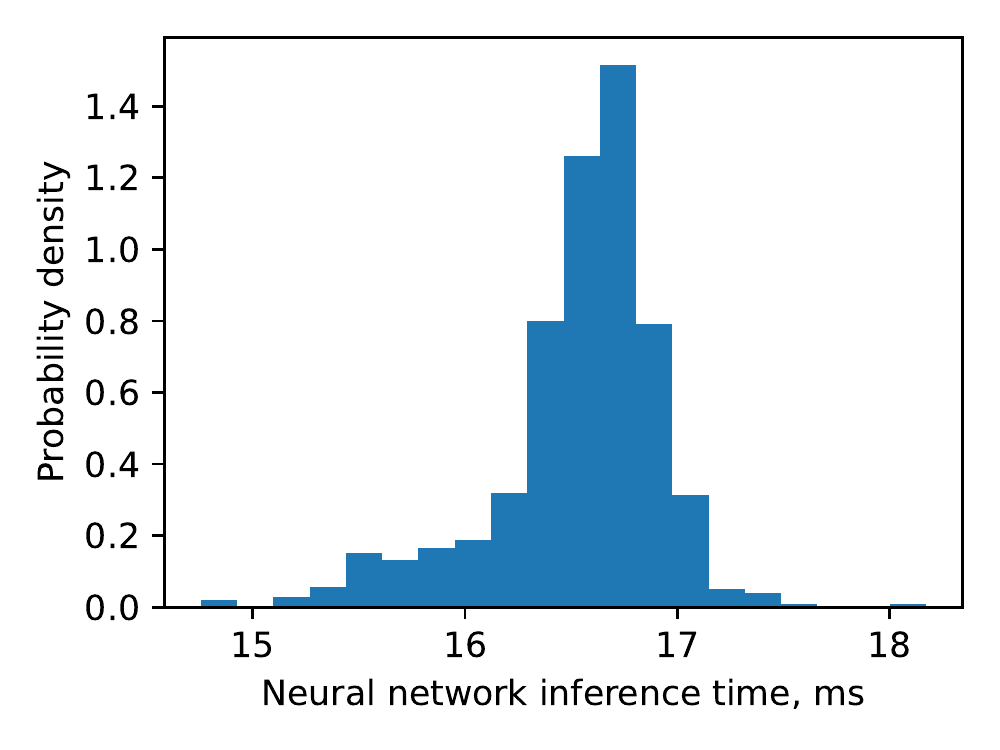}
    \caption{Empirical distribution of neural network inference time.}
    \label{fig:nn_distr}
\end{figure}

The flow graph nodes' execution times are sampled from empirical distributions measured by profiling the real system.
Fig.~\ref{fig:nn_distr} contains an example of such distribution for the neural network inferencer.
The oneTBB flow graph nodes are profiled using Intel Flow Graph Tracer and Flow Graph Analyzer \citep{Voss2019flowgraphbeyond}.
The non-oneTBB activities like the video decoding and the neural network inference are sampled by our in-house tracing library. 

We deliberately use flow graphs in the model instead of some well-known formalisms like Petri Nets \citep{peterson1977petri} and Queuing Networks \citep{chandy1975parametric}. 
It makes designing the simulation model easier because we can explicitly use the same graph structure and synchronization primitives used as the real system.
This approach also has potential for the model to be automatically generated by parsing the profiling traces.
Moreover, our approach is easier to be adopted by the software developers who are unlikely to be experts in modelling and simulation.

We shall notice that we do not consider the overhead caused by communications between the nodes, as the execution time of each node significantly exceeds the average communication time in our case.
However, our simulation model can easily be upgraded by introducing additional timings between consecutive graph nodes.
This will allow to model other AI applications like AI on pure-cloud solutions, where there is significant communication overhead cause by the network.
The communication overhead can also be important in self-driving cars, where the TCP/IP stack is often used for communication even within a single machine, e.g. the ROS framework\citep{quigley2009ros}.

\section*{\textbf{EVALUATION}}
\label{sec:evaluation}
In this section we evaluate the proposed simulation model.
In order to do it, we compared the performance metrics predicted by the simulator with those measured on the real video analytics system. 
In our case, the performance metrics is the average frames per second (FPS) per each video stream. 

The hardware specifications are listed in Table~\ref{tb:hardware_specs}.
\begin{table}[h]
    \centering
    \caption{Testbed hardware specifications}
    \label{tb:hardware_specs}
    \begin{tabular}{|l|l|l|}
    \hline
        Parameter & Value \\ \hline
        CPU & Intel Core i7-7800X, 3.5 GHz, \\
            & 6 Cores, 12 Threads \\ \hline
        GPU & NVIDIA GeForce GTX 1080 Ti, \\ 
            & 11 GB VRAM \\ \hline
        RAM & 64 GB \\ \hline
    \end{tabular}
\end{table}

\subsection*{\textbf{Estimating the model accuracy and the system scalability}}
\begin{figure}[t!]
    \centering
    \includegraphics[width=1\linewidth]{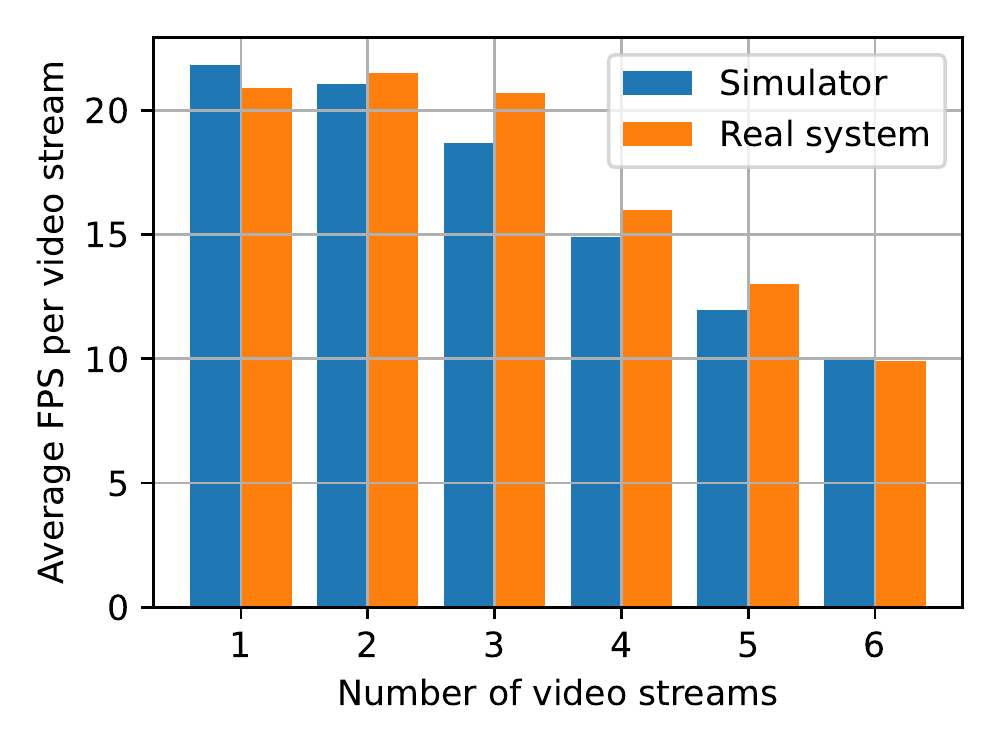}
    \caption{FPS per video stream depending on the number of video streams.}
    \label{fig:evaluation}
\end{figure}

In this experiment, we varied the number of video streams in order to estimate the accuracy of the simulation model.
The results presented in Fig.~\ref{fig:evaluation} demonstrate, that the performance prediction error rate of the simulation model is less than 10\%, which is accurate enough to rely on the model's prediction in planning the scalability of the system.
It is also lower than the error rate threshold of 20\% used in a related work \citep{bian2014simulating}.

On small number of video streams (up to three) the system experiences almost no performance drop due to efficient parallelism.
However, when the number of video streams increases, the performance drops significantly, because the neural network becomes the bottleneck that hinders the parallelism: threads stand idle waiting for a task to execute. 
When the number of video streams exceeds the number of CPU logical cores, the performance drops even more because there is not enough free threads to execute appearing tasks. 



\subsection*{\textbf{Integration of a new module in the system}}
\begin{figure}[t!]
    \centering
    \includegraphics[width=1\linewidth]{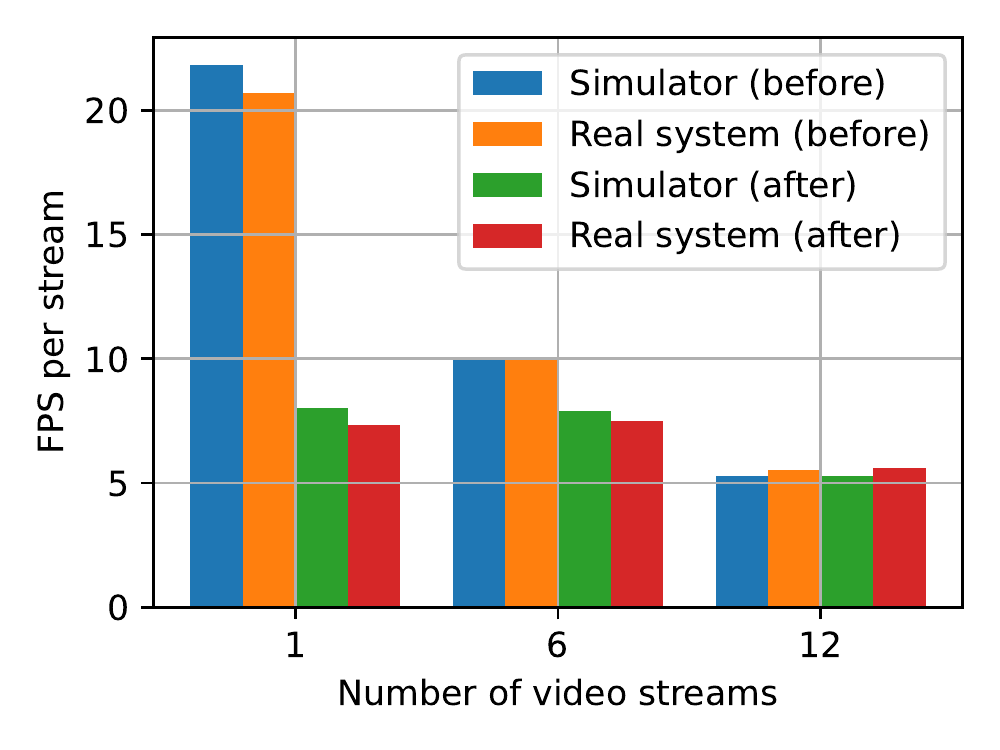}
    \caption{Performance impact caused by integrating a new CPU-intensive module}
    \label{fig:slowdown}
\end{figure}

To study an impact of a new module on the system, we conducted the following experiment.
We added a CPU-intensive module (\texttt{Node E} in Fig~\ref{fig:vas_graph}) in the system and measured the overall slowdown on both the real model and the simulator.
The results, see Fig~\ref{fig:slowdown}, show that the module produces approximately 2.5X slowdown on 1 video stream, meanwhile producing no noticeable slowdown on 12 video streams.
The result is counter-intuitive: the increase in workload makes the overhead of the module unnoticeable instead of it growing linearly with the workload.
This occurs because the neural network, while being comparably fast on one stream, becomes the bottleneck on high number of video streams.

This experiments demonstrates, that for systems with high degree of parallelism the impact of adding a new module or changing the existing one on system's performance could be nontrivial.

\subsection*{\textbf{Estimating the feasibility of software optimizations}}
In this section we study the feasibility of optimizations in the considered system with the help of the proposed simulation model by the example of the following optimization.

The quality of service (QoS) of the considered video analytics system highly depends on tracking algorithms.
Tracking accuracy, in turn, depends on many parameters of the algorithm.
In order to improve QoS, it is needed to find the optimal values of the parameters.

The simplest way to find the optimal values is the grid search.
However, for the high-dimensional search space of parameters it is computationally unfeasible.
More sophisticated optimization methods like Bayesian optimization allow to mitigate but not fully alleviate this problem, still requiring many runs of the system.

Fortunately, all the required runs use the same input data with different values of parameters.
Therefore, in theory, it is possible to save a lot of time by caching neural network predictions and reusing them on each run instead of inferring the real network.

However, the cache may not provide the desired performance gain, because some other module can become the bottleneck.
Therefore, prior to implementing and testing the optimization (which takes about one week for a software engineer), we study the feasibility of the optimization on the simulation model (which takes a couple of hours).

First, we used the developed simulation model to estimate the performance gain of implementing the neural network cache without considering the overhead of the cache itself.
Effectively, we set the execution time of the GPU node to zero.
The speed-up appeared to be 13.8x (Table~\ref{tbl:nn_cache}, ``ideal'' cache).

Then we implemented a LevelDB-based\footnote{More details at: \url{https://github.com/google/leveldb}} cache module detached from the system, benchmarked its performance and used the data in simulation model to account for the overhead.
The speed-up appeared to be 12.0x (Table~\ref{tbl:nn_cache}, ``real'' cache).

Finally, we integrated the developed cache module into the system.
The real achieved speed-up was equal to 11.3x (Table~\ref{tbl:nn_cache}, real system), close to the predicted value of 12.0x.

The experiment demonstrates, that with a negligible overhead in time spent on simulating each step of the implementation, it allows to correctly estimate the limit of optimization's impact.
Moreover, it could save a lot of time, if it had emerged that a specific optimization step is not worth implementing.

\begin{table}[]
    \centering
    \caption{Overall system speedup on 6 video streams from implementing the cache}
    \label{tbl:nn_cache}
    \begin{tabular}{|l|l|}
        \hline
        Experiment               & Overall system speedup \\ \hline
        Real system              & 11.3x                   \\ \hline
        Simulator, ``ideal'' cache & 13.8x                   \\ \hline
        Simulator, ``real'' cache  & 12.0x                   \\ \hline
    \end{tabular}
\end{table}
\section*{\textbf{CONCLUSIONS}}
In this work, we proposed a discrete-simulation model to predict the performance of a heterogeneous CPU/GPU video analytics system.
The proposed model can easily be adopted by the software developers who are not experts in simulation and modelling.
We showed that the accuracy of performance estimation using the proposed system is higher than 90\% in each experiment. 

We used the simulation model to predict performance scale with workload; to estimate the impact of a new module on the whole system, which demonstrated counter-intuitive results yet correctly predicted by the simulator; to predict the feasibility of an optimization.

We believe such a simulation model should become a workplace tool for software designers and it could save lots of resources by easily inferring the feasibility of optimizations and modifications prior to doing some costly work on a real system. 

Possible future research includes taking into consideration the hardware configuration to predict, for example, an optimal hardware price --- quality of service trade-off of a system.

 
\renewcommand{\bibsection}{}
\setlength{\bibhang}{18pt}
\section*{\textbf{REFERENCES}}
\bibliographystyle{ecms}
\bibliography{main}

\vspace{0.75\textheight}

\section*{\textbf{AUTHOR BIOGRAPHIES}}

\begin{wrapfigure}{l}{0.085\textwidth}
    \includegraphics[width=0.135\textwidth]{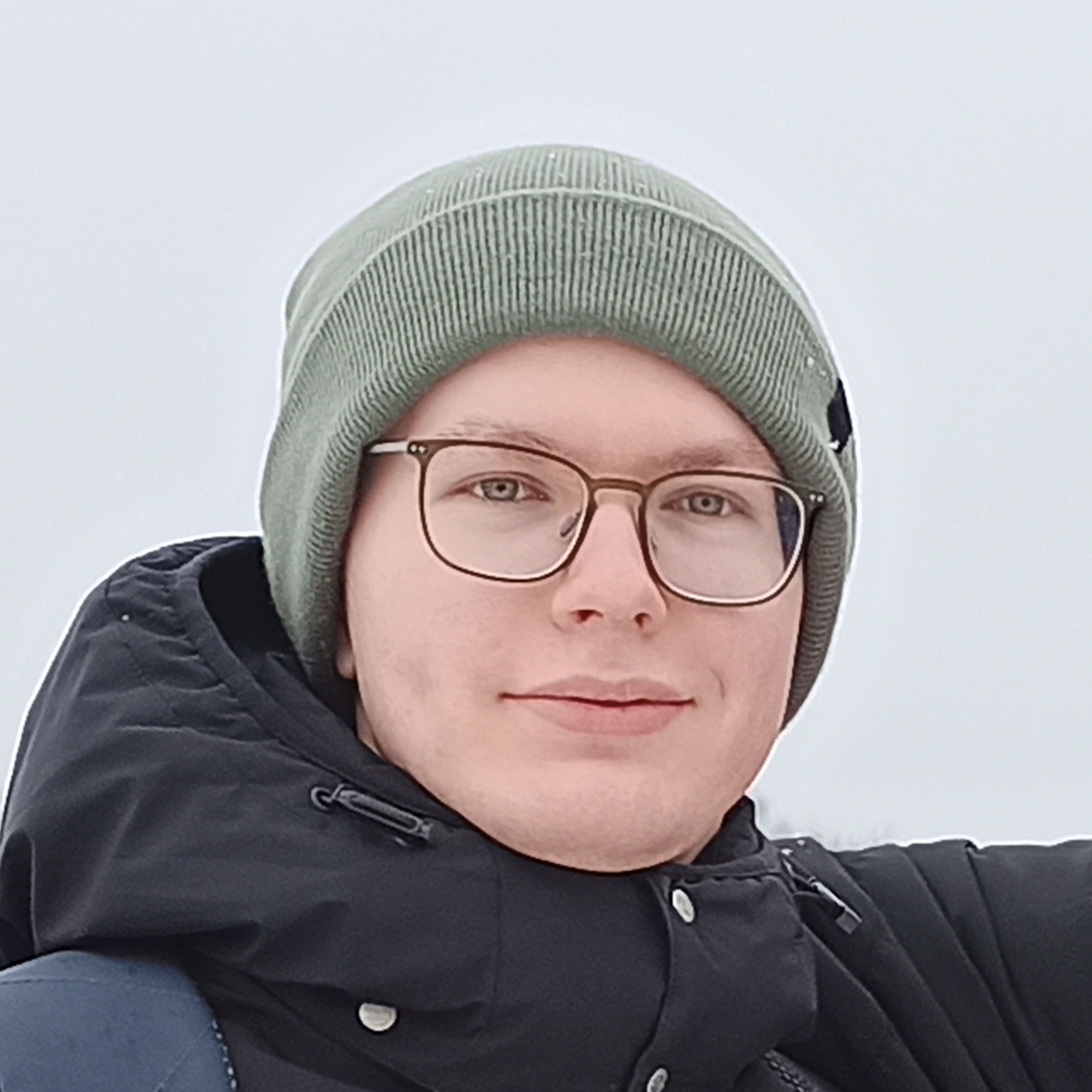}
\end{wrapfigure}
\noindent \textbf{{\MakeUppercase{Vyacheslav Zhdanovskiy}}} 
\noindent  was born in Verkhnyaya Pyshma, Russia. 
He obtained his B.Sc. in Applied Physics and Mathematics in 2020 from Moscow Institute of Physics and Technology (MIPT). 
Currently he is a M.Sc. student in Computer Science and Engineering at MIPT.
Since 2020, he works at the Vision Systems Lab at the Institute for Information Transmission Problems.
His research interests include heterogeneous and parallel computing, computer vision and distributed systems. His e-mail address is zhdanovskiy.vd@phystech.edu.
\\

\begin{wrapfigure}{l}{0.1\textwidth}
  \includegraphics[width=0.14\textwidth]{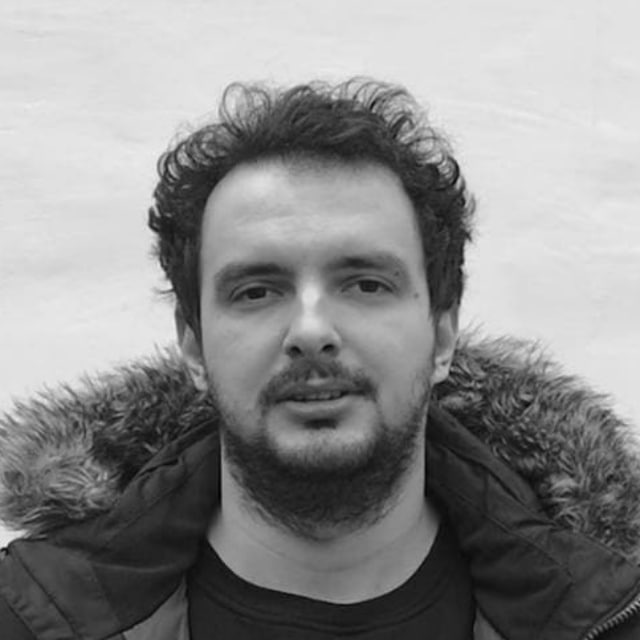}
\end{wrapfigure}
\noindent \textbf{{\MakeUppercase{Lev Teplyakov}}} 
\noindent was born in Arkhangelsk, Russia.
He obtained his B.Sc. and M.Sc. in Applied Physics and Mathematics from Moscow Institute of Physics and Technology (MIPT) in 2017 and 2019 correspondingly.
Since 2016, he has been developing industrial computer vision systems with the Vision Systems Lab at the Institute for Information Transmission Problems.
His research interests include heterogeneous and parallel computing, object detection and tracking.
His e-mail address is teplyakov@visillect.com.
\\

\begin{wrapfigure}{l}{0.1\textwidth}
  \includegraphics[width=0.14\textwidth]{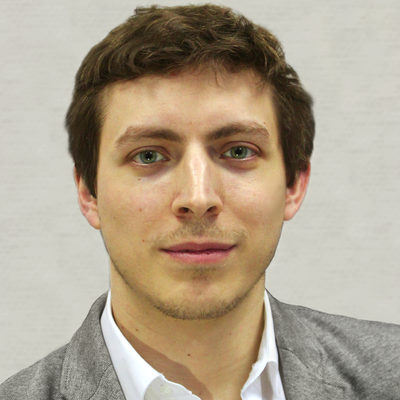}
\end{wrapfigure}
\noindent \textbf{{\MakeUppercase{Anton Grigoryev}}} 
\noindent was born in Petropavlovsk-Kamchatskiy, Russia.
Having graduated from Moscow Institute of Physics and Technology, he has been developing industrial computer vision systems with the Vision Systems Lab at the Institute for Information Transmission Problems since 2010.
His research interests are image processing and enhancement methods, autonomous robotics and software architecture.
His e-mail address is me@ansgri.com.

\end{document}